\documentclass[a4paper,amsmath,preprint,
               superscriptaddress,nofootinbib,floatfix]{revtex4}

\usepackage{graphicx}

\newcommand{\bb}{\ensuremath{B^0-\bar{B}^0}}
\newcommand{\kk}{\ensuremath{K^0-\bar{K}^0}}

\newcommand{\acp}{\ensuremath{A_{\rm CP}^{\rm mix}(B\to J/\psi\,K_S)}}
\newcommand{\ambsg}{\ensuremath{A_{\rm CP}^{\rm mix}(B\to M_s\,\gamma)}}
\newcommand{\adbsg}{\ensuremath{A_{\rm CP}^{\rm dir}(B\to X_s\,\gamma)}}

\newcommand{\bsg}{\ensuremath{b\to s\,\gamma}}
\newcommand{\meg}{\ensuremath{\mu\to e\,\gamma}}
\newcommand{\ek}{\ensuremath{\varepsilon_K}}
\newcommand{\dmbd}{\ensuremath{\Delta m_{B_d}}}
\newcommand{\dmbs}{\ensuremath{\Delta m_{B_s}}}
\newcommand{\dmbsd}{\ensuremath{\Delta m_{B_s}/\Delta m_{B_d}}}

\begin{document}
\title{Exploring Flavor Structure of Supersymmetry Breaking\\
       at $B$ factories
\footnote{Talk given by T.~Shindou at the 3rd Workshop on Higher Luminosity 
$B$ Factory, August 6--7, 2002, Shonan Village, Japan.}}


\author{Toru Goto}
\email{goto@het.phys.sci.osaka-u.ac.jp}
\affiliation{Department of Physics,
Graduate School of Science,  Osaka University, Toyonaka, Osaka 560-0043, Japan}

\author{Yasuhiro Okada}
\email{yasuhiro.okada@kek.jp}
\affiliation{Theory Group, KEK, Tsukuba, Ibaraki 305-0801, Japan}
\affiliation{Department of Particle and Nuclear Physics,
             The Graduate University of Advanced Studies,
             Tsukuba, Ibaraki 305-0801, Japan}

\author{Yasuhiro~Shimizu}
\email{shimizu@eken.phys.nagoya-u.ac.jp}
\affiliation{Department of Physics, Nagoya University, Nagoya 464-8602, Japan}

\author{Tetsuo Shindou}
\email{shindou@het.phys.sci.osaka-u.ac.jp}
\author{Minoru Tanaka}
\email{tanaka@phys.sci.osaka-u.ac.jp}
\affiliation{Department of Physics, Graduate School of Science,
             Osaka University, Toyonaka, Osaka 560-0043, Japan}
\begin{abstract}
We investigate flavor physics at present and future $B$ factories
in order to distinguish supersymmetric models.
We evaluate CP asymmetries in various $B$ decay modes,
$\dmbd$, $\dmbs$, and $\ek$ in three supersymmetric models, 
i.e. the minimal supergravity, 
the SU(5) SUSY GUT with right handed neutrinos, and 
a supersymmetric model with U(2) flavor symmetry.
The allowed regions of $\dmbsd$ and
CP asymmetries in $B\to J/\psi\,K_S$  and $\bsg$ are different for
the three models so that it is possible to distinguish the three models
by precise determinations of these observables in near future experiments.
\end{abstract}
\preprint{hep-ph/0211143}
\preprint{OU-HET 425}
\preprint{KEK-TH-854}
\preprint{DPNU-02-36}
\maketitle
\section{Introduction}
CP violation in $B$ system has been established
by measurements of Belle experiment at KEK\cite{belle} 
and BaBar experiment 
at SLAC\cite{babar}.
In the standard model(SM), the phenomenon of CP violation originates
from Kobayashi-Maskawa mechanism\cite{KM} and the experimental 
results are 
consistent with the predictions of the SM. In near future, it is expected
that measurements of CP violation and rare decay processes
will be improved at the asymmetric $B$ factories and that the magnitude
of $B_s^0$--$\bar{B}_s^0$ mixing will be determined at the Fermilab Tevatron 
experiments\cite{tevatron}. Within a few years, we will be able to 
know whether or not significant effects of new physics reside
in the $B$ systems.

Supersymmetry(SUSY) is one of the most attractive candidates of
physics beyond the SM. SUSY 
gives a justification to the electroweak scale, which is much smaller 
than the Planck scale. In SUSY models, there exist some features:
(i) the gauge coupling constants can be unified, (ii) the existence
of light Higgs boson is predicted, (iii) there are superpartners of
the SM particles, and so on.

Mass matrices of superpartners of quarks and leptons
are new sources of flavor mixing and CP violation. 
These mass matrices are determined from
SUSY breaking terms in the Lagrangian, and these terms reflect the
SUSY breaking mechanism and interactions that present between the scale
of the SUSY breaking and the electroweak scale.
For general soft SUSY breaking, if the squark and slepton masses are
below a few TeV,
the experimental data of $\ek$ which is related to the CP violation
in $\kk$ mixing and $\mu\to e\gamma$ require 
strong degeneracy between the first- and second-family squarks and sleptons.
In order to realize this degeneracy, many flavor models are considered 
and these models predict different flavor structures.
Therefore if SUSY particles are discovered at
LHC and/or a linear collider experiment, 
$B$ physics is expected to play an important role
in studying the flavor structure of SUSY models.

Now we address the following question: Can we distinguish 
SUSY models by measurements at $B$ factories?
In order to answer this question, we investigate SUSY effects on 
$B$ physics in three different SUSY models, namely (i) the minimal
supergravity(mSUGRA) model, (ii) the SU(5) SUSY grand unified 
theory (GUT) with right-handed neutrinos\cite{su5}, and 
(iii) a supersymmetric model with U(2) flavor
symmetry\cite{u2_1,u2_2}.
In our recent paper\cite{gosst}, we focus on the new physics search through 
the consistency check of the unitarity triangle and
we evaluate SUSY effects on CP asymmetries in 
$B\to J/\psi K_S$, $\dmbs$, $\dmbd$, $\ek$, and 
$\phi_3\equiv \arg(-V_{ub}^*V_{ud}/V_{cb}^*V_{cd})$. 
In this talk, besides these
observables, we also evaluate both direct and mixing CP asymmetries in 
$B\to X_s\gamma$.

\section{Models}

Here, we give a brief description of models.
A detailed discussion of these models can be found in Ref.~\cite{gosst}.

\subsection{The minimal supergravity model}

In the mSUGRA, SUSY is spontaneously broken in the hidden 
sector and the hidden sector of the theory communicates with our
MSSM sector only through the gravitational interaction.
The soft breaking terms are induced through the gravitational 
interaction and the structure of soft breaking
terms have no new flavor mixing
at the scale where the soft breaking terms are induced.

In this model, the source of flavor mixing is only
CKM mixing matrix.
The universality of the squark sector is lost due to radiative 
corrections induced by the CKM mixing.
We use renormalization group(RG) equations in order
to trace these radiative corrections and determine the soft breaking 
terms at the electroweak scale.

\subsection{The SU(5) SUSY GUT with right-handed neutrinos}

The measurements of three gauge coupling constants support
the idea of the supersymmetric grand unification. Furthermore,
recent results of neutrino oscillation experiments suggest the
existence of finite mass of neutrinos. In order to explain these
points, the SU(5) SUSY GUT with right-handed neutrinos is 
considered\cite{su5}.

In this model, the soft breaking terms are the same as
in the mSUGRA model at the scale where the soft breaking
terms are induced.
Unlike the mSUGRA model, the SU(5) SUSY GUT with 
right-handed neutrinos have two sources of flavor mixing.
One is CKM matrix as in the case of the mSUGRA.
Another is Maki-Nakagawa-Sakata matrix\cite{mns}, which is 
the mixing matrix of the lepton sector. 
A large flavor mixing in the neutrino sector
can affect the squark mixing in the right-handed down type
sector through GUT interactions.

\subsection{A model with U(2) flavor symmetry}

Instead of assuming the universality of sfermion mass matrices,
it is possible to solve the flavor problem of SUSY by 
introducing some flavor symmetry. U(2) flavor symmetry is one of
such symmetries. We use the model given in Ref.~\cite{u2_2}.

In this model, the quark and lepton supermultiplets in the first
and the second generations transform as doublets under the U(2)
flavor symmetry while the third generation and the Higgs supermultiplets
are singlet under the U(2).

In order to obtain the correct structure of Yukawa couplings, we assume
that the breaking pattern of the U(2) is
\begin{equation}
\text{U(2)}\longrightarrow\text{U(1)}\longrightarrow\openone
(\text{no symmetry}).
\label{U2breaking}
\end{equation}
With this U(2) breaking, we obtain the quark Yukawa couplings $f_Q$
and the squark mass matrices $m_X^2$:
\begin{equation}
(f_Q^{ij})=y_Q \left(\begin{array}{ccc}
                      0               & a_Q\,\epsilon' & 0          \\
                      -a_Q\,\epsilon' & d_Q\,\epsilon  & b_Q\,\epsilon\\
                      0               & c_Q\,\epsilon  & 1          \\
                     \end{array}\right),\quad Q=U,D,
\label{U2Yukawa}
\end{equation}
\begin{equation}
m_X^2=m_0^{X2}\left(\begin{array}{ccc}
                    1 & 0 & 0 \\
                    0 & 1+r_{22}^X\epsilon^2 & r_{23}^{X}\epsilon \\
                    0 & r_{23}^{X*}\epsilon & r_{33}^{X}
                    \end{array}
              \right),\quad X=Q,U,D,
\label{U2smassmatrix}
\end{equation}
where $\epsilon\simeq \lambda^2$ and 
$\epsilon^{\prime}\simeq \lambda^3$ are parameters of the U(2) and U(1)
symmetry breakings respectively,
and
$y_Q$'s, $a_Q$'s, $b_Q$'s, $c_Q$'s, $d_Q$'s, and $r^X$'s
are dimensionless constant parameters of $O(1)$.

In the mass matrices of sfermions in this model, 
the degeneracy between masses of the first and the second 
generation is naturally realized, while mass of the third generation
may be separated from the others and there exist flavor mixing
between the 2nd and the 3rd generation of sfermions.
This is new source of flavor mixing besides CKM mixing matrix.

\section{Numerical Analysis}
\subsection{Parameters}
The parameters and experimental constraints used in our 
calculation are the same
as those in Ref.~\cite{gosst} except for the analysis of CP asymmetries
in $\bsg$. In the analysis of CP asymmetries in $\bsg$,
we introduce new CP violating phases on 
the tri-linear scalar couplings($A$-terms).

\subsection{Numerical results}

\begin{figure}
\includegraphics[scale=0.8]{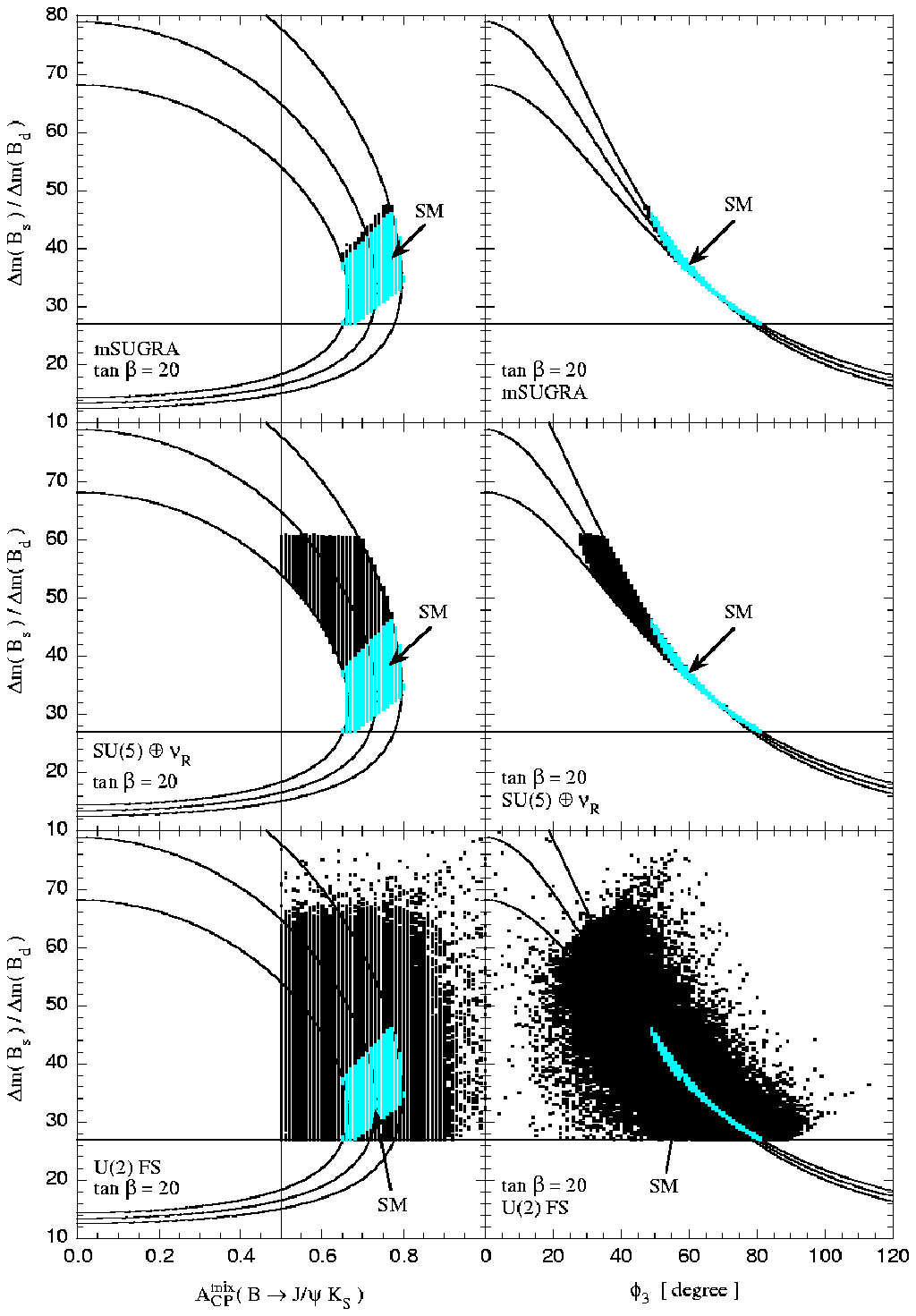}
\caption{
Scatter plots in the planes $(\acp,\,\dmbsd)$ and $(\phi_3,\,\dmbsd)$
for three SUSY models.
Solid curves show the SM values with fixed $|V_{ub}/V_{cb}|=0.08$, $0.09$
and $0.10$. 
}
\label{fig:dmbsd-acp-phi3}
\end{figure}

First, we discuss SUSY contributions to 
$\dmbsd$, $\acp$, \ek, and $\phi_3$. We vary $|V_{ub}/V_{cb}|$ 
and $\phi_3$ and impose experimental constraints such as
$B(\bsg)$, $B(\meg)$, the measured values of $\ek$ and $\dmbd$, and
the lower bound of $\dmbs$. 
In Fig.~\ref{fig:dmbsd-acp-phi3}, 
we show possible values of $\acp$, $\dmbsd$, and $\phi_3$ for 
the parameter sets  which satisfy the constraints.

In the mSUGRA, the deviations from the SM values are not 
significant. Because there are at most 10\% deviations in 
$\ek$ while $\dmbs$, $\dmbd$ and $\acp$ are almost the same as
the values of the SM,
it seems that there are at most 10\% deviations in both
the $(\acp, \dmbsd)$ plane and the $(\acp, \phi_3)$ plane.

In the SU(5) SUSY GUT with right-handed neutrinos, though
we can see the deviation of order one in each observable, 
all the allowed points lie between the lines corresponding to
the SM values with $|V_{ub}/V_{cb}|=0.08$ and $0.10$.
This is due to the fact that there exist significant SUSY contribution
only to the $\kk$ mixing, not to the $\bb$ mixing.
There can be SUSY contribution of order one to $\ek$, 
while $\dmbs$, $\dmbd$, and $\acp$ are not affected by
SUSY.

In the U(2) model, 
there are SUSY contributions to both the $\kk$ mixing and the $\bb$ mixing.
In addition to the huge SUSY contribution to
$\ek$,
there are correction of order one to $\dmbs$ and $\dmbd$.
Therefore the $\dmbsd$, $\acp$, and $\phi_3$ can be 
considerably different from the
SM values.

\begin{figure}
\includegraphics{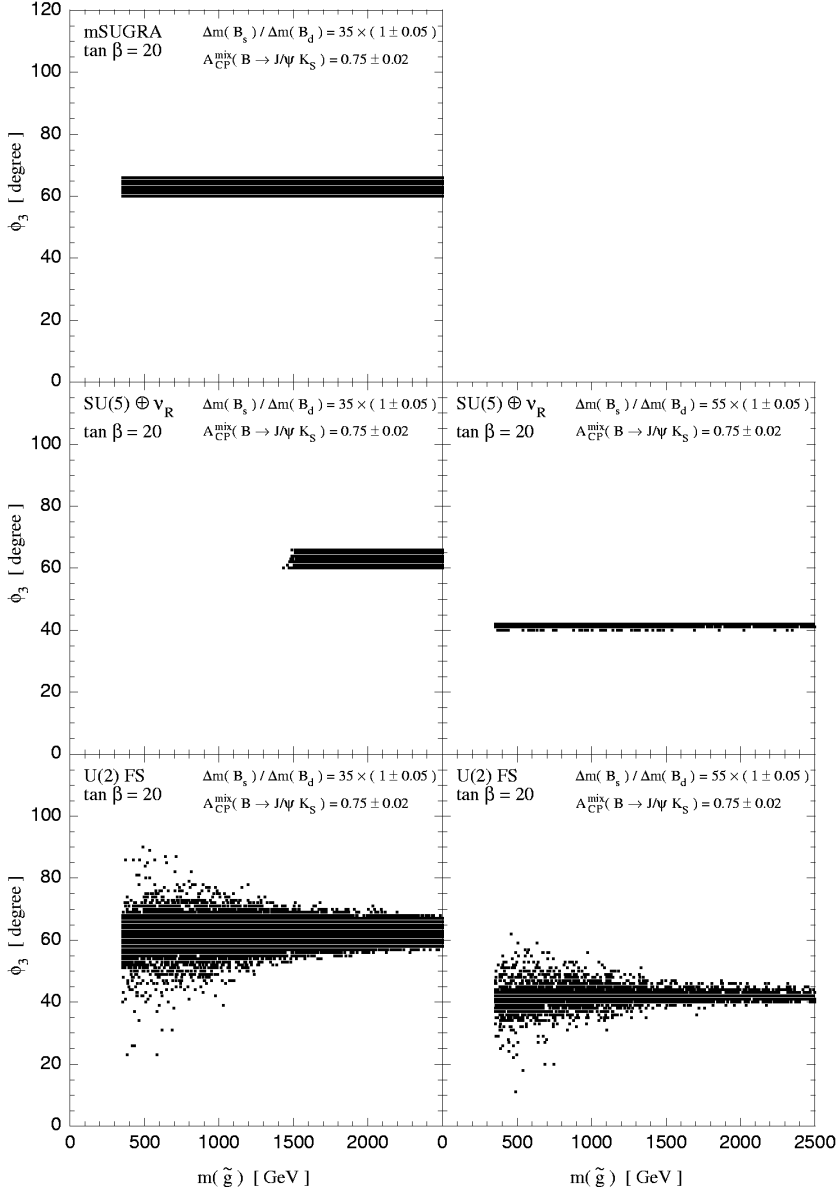}
\caption{
Possible regions of $\phi_3$ as a function of the gluino mass.
}
\label{fig:phi3-gno}
\end{figure}

Here we discuss future prospects of new physics search in $B$ physics.
It is expected that $\acp$ and $\dmbsd$ will be precisely measured in near
future at the $B$ factories and Tevatron experiments. As an illustration,
we consider two cases that $\dmbsd$ and $\acp$ are precisely determined
at future $B$ experiments as
\begin{displaymath}
\begin{array}{cll}
 \mbox{(a)} & \dmbsd = 35\times(1\pm0.05), & \acp = 0.75\pm0.02,
\\
 \mbox{(b)} & \dmbsd = 55\times(1\pm0.05), & \acp = 0.75\pm0.02.
\end{array}
\end{displaymath}
(a) corresponds to the case where $\acp$, and $\dmbsd$ are consistent
with the SM.  (b) is the case where there is some inconsistency
among the three observables within the SM.
In Fig.~\ref{fig:phi3-gno}, we present the possible region of 
$\phi_3$ in the cases (a) and (b) for the three models.

For the case (a), $\phi_3=60^{\circ}-65^{\circ}$ in the SM.
The values of $\phi_3$ in the mSUGRA and the SU(5) SUSY GUT
with right-handed neutrinos are similar to the value of the SM.
On the other hand, in the U(2) model, the $\phi_3$ value 
may be different from the value in the SM.

For the case (b), as well as the SM, the mSUGRA is excluded.
In the other two models, there exist the allowed regions of $\phi_3$
value.

Secondly, we evaluate the direct CP asymmetry $\adbsg$
in the inclusive decays $B\to X_s\gamma$ and the mixing
induced CP asymmetry $\ambsg$ in the exclusive decays 
$B^0\to M_s \gamma$ where $M_s$ denotes a CP eigen state
which includes a strange quark such as $K^*$ and $K_1$. 
It is expected that $\adbsg$ would be a clean signal of 
new physics\cite{adbsg_1,adbsg_2},
because in the SM it can be calculated precisely and is found to 
be about 0.5\%\cite{adbsg_2}. 
Though the value of $\ambsg$ in the SM is at a level of 1\%\cite{ambsg_1},
this value may significantly change
with new physics contributions\cite{ambsg_2}.

\begin{figure}
\includegraphics{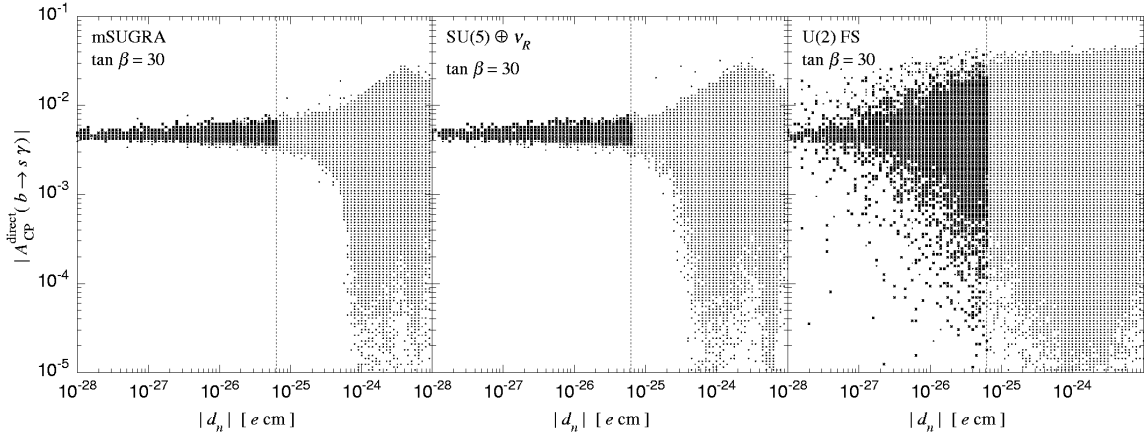}
\caption{
Scatter plots in the plane of $\adbsg$ and the electric dipole moment
of the neutron $|d_n|$ for three SUSY models. The gray points are
excluded by the experimental constraint of $|d_n|$.
}
\label{fig:absg-edm}

\end{figure}

In Fig.~\ref{fig:absg-edm}, we show possible regions of $\adbsg$ and
the electric dipole moment(EDM) of the neutron with
its experimental bound $|d_n|<0.63\times 10^{-25}~e\cdot \text{cm}$. 
The values of EDM of the neutron become larger 
as the deviation of $\adbsg$ increases.
This constraint of EDM is so strong that 
in the mSUGRA and the SU(5) SUSY
GUT with right-handed neutrinos, the value of $\adbsg$ is about
0.5\% which is predicted in the SM for the allowed value of EDM of 
the neutron.
In the U(2) model,
$\adbsg$ can be as large as a few percents even under the constraint
from the EDM.

\begin{figure}
\includegraphics{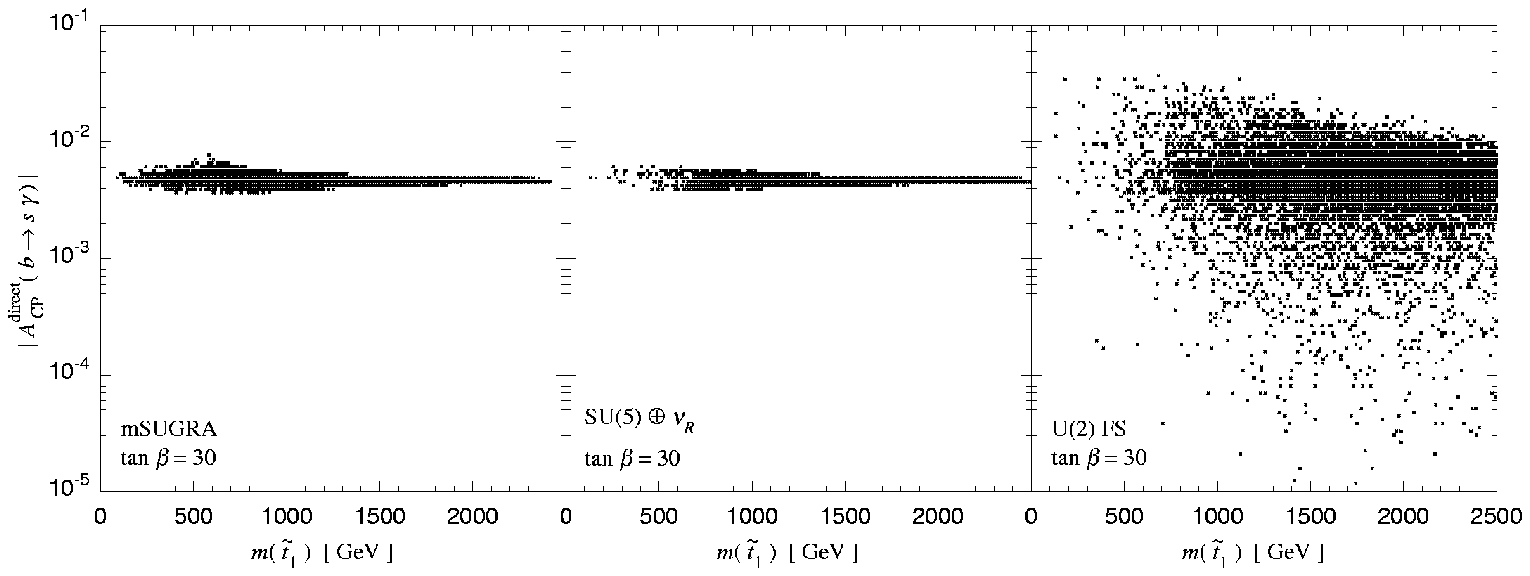}
\caption{
Possible regions of $\adbsg$ as a function of the lightest stop mass.
}
\label{fig:absg-stp}
\end{figure}

The allowed region of $\adbsg$ is shown as a function of the 
lightest stop mass in Fig.~\ref{fig:absg-stp}.
In the U(2) model, $\adbsg$ can be as large as a few percent
even if the mass of the lightest stop is as large as 2.5TeV,
while in the mSUGRA and the SU(5) SUSY GUT with right-handed
neutrinos the value of $\adbsg$ is about 0.5\% in all range of
the lightest stop mass.

\begin{figure}
\includegraphics{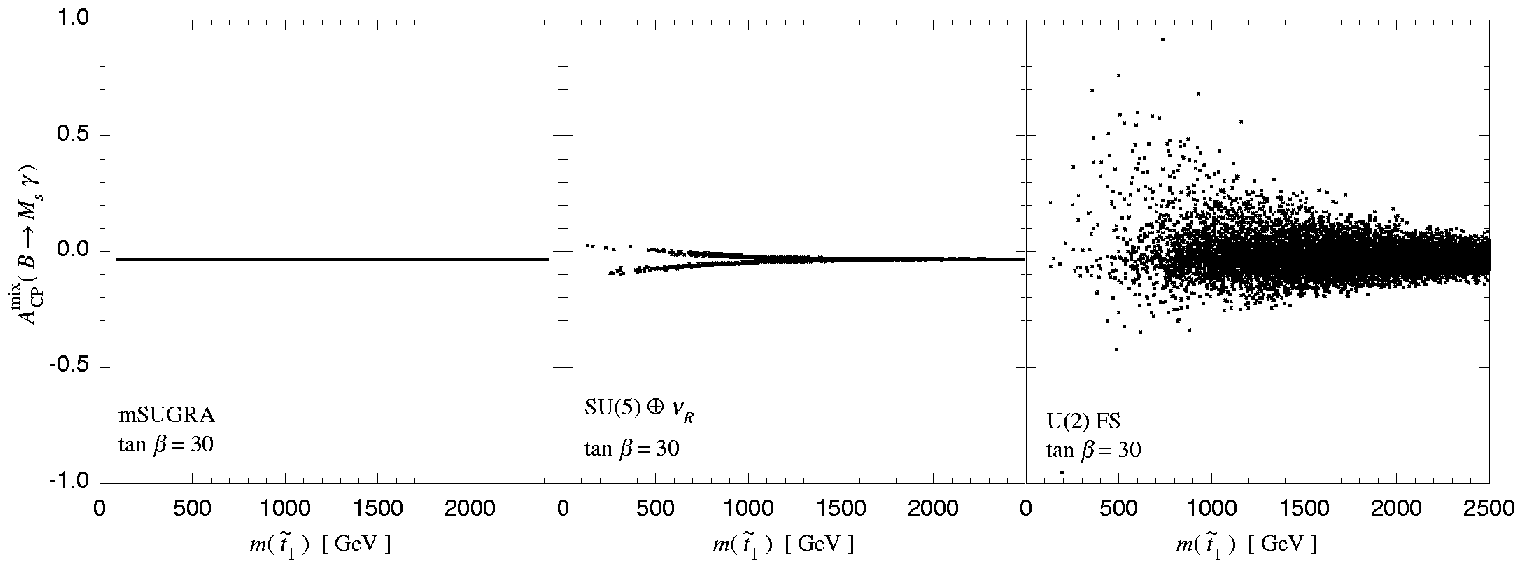}
\caption{
Possible region of $\ambsg$ as a function of the lightest stop mass.
}
\label{fig:ambsg-stp}
\end{figure}

In Fig.~\ref{fig:ambsg-stp}, the possible regions of $\ambsg$ are
plotted as a function of the lightest stop mass. In the mSUGRA,
$\ambsg$ is almost the same as the value of the SM.
In the SU(5) SUSY GUT with right-handed neutrinos,
for the region where the lightest stop mass $m(\tilde{t}_1)$ is
smaller than $1\text{TeV}$, we find that $\ambsg$ is different
from the value of the SM and becomes at most about 10\%.
In the U(2) model, $\ambsg$
can be as large as 0.5 for the range of $m(\tilde{t}_1)<1\text{TeV}$.
Even in the case of $m(\tilde{t}_1)>1\text{TeV}$, $\ambsg$ in the 
U(2) model can be as large as 0.2.

\section{Summary}

In order to distinguish SUSY models by measurements at $B$ factories,
we have evaluated SUSY contributions to  several observables in $B$ physics.
We use three typical SUSY models, namely, mSUGRA, the SU(5) SUSY GUT with
right-handed neutrinos, and the U(2) model.

First we have considered $\dmbs$, $\dmbd$, $\acp$, 
$\ek$, and $\phi_3$.
In the mSUGRA, the deviations from the SM values exist only in 
$\ek$ and is at most 10\% and $\dmbs$, $\dmbd$, $\acp$, and $\phi_3$
are almost the same as in the SM.
In the SU(5) SUSY GUT with right-handed neutrinos,
there is deviation of order one in $\ek$ from the value of the SM.
As well as in the mSUGRA, $\dmbs$, $\dmbd$, $\acp$, and $\phi_3$
are not affected by SUSY contributions.
In the U(2) model, there are so large contribution to $\ek$
that the value of $\ek$ can be as ten times large as
the value of the SM.
In addition, the SUSY contributions affect not only 
$\ek$ but also $\dmbs$, $\dmbd$, $\acp$, and $\phi_3$.
The deviations of $\dmbs$, $\dmbd$ from the values
of the SM can be order one.

In the case that the two observables $\dmbsd$ and $\acp$ are precisely 
determined at the $B$ factories in near future, 
it may be possible to distinguish the different
SUSY models by measuring $\phi_3$ at the $B$ factories
and Tevatron experiments.

Secondly, we have investigated SUSY contributions to $\adbsg$ and 
$\ambsg$. Owing to the constraint from the EDM of the neutron,
the values of $\adbsg$ in the mSUGRA and the SU(5) SUSY GUT with
right-handed neutrinos are almost consistent with the value expected
in the SM.
On the other hand, in the U(2) model, $\adbsg$ can be as large
as a few percent which is ten times larger than the value in the SM. 

In the mSUGRA, $\ambsg$ is almost the same as in the SM for all
range of the lightest stop mass.
In the SU(5) SUSY GUT with 
right-handed neutrinos, $\ambsg$ can be as large as 10\% if
the lightest stop mass is smaller than 1TeV.
We have shown that 
in the U(2) model, $\ambsg$ can be $\approx 0.5$ in the parameter
region where the lightest stop mass is smaller than 1TeV.
Even in the case that the lightest stop mass is larger than
1TeV, a 20\% asymmetry in $B\to M_s\gamma$ is possible.

In conclusion,
as we have illustrated with three specific models, 
SUSY models with different flavor structures exhibit 
different patterns of the deviations from the SM in the $B$ physics.
Therefore experiments
at $e^+e^-$ $B$ factories and hadron machines are very important to explore the
flavor structure of SUSY breaking.

\end{document}